\documentclass[aps,prl,reprint,superscriptaddress]{revtex4-1}
\usepackage{graphicx}
\usepackage{color}
\usepackage{amsmath}
\begin{document}

\title{Effect of Shape and Friction on the Packing and Flow of Granular Materials}

\author{K. Michael Salerno}
\affiliation{US Army Research Laboratory, Aberdeen Proving Ground, MD 21005, USA}
\author{Dan S. Bolintineanu}
\author{Gary S. Grest}
\author{Jeremy B. Lechman}
\author{Steven J. Plimpton}
\author{Ishan Srivastava}
\affiliation{Sandia National Laboratories, Albuquerque, New Mexico 87185, USA}
\author{Leonardo E. Silbert}
\affiliation{School of Math, Science, and Engineering, Central New Mexico Community College, Albuquerque, New Mexico 87106, USA}

\begin{abstract}

The packing and flow of aspherical frictional particles are studied using
discrete element simulations. Particles are superballs with shape
$|x|^{s}+|y|^{s}+|z|^{s} = 1$ that varies from sphere ($s=2$) to cube
($s=\infty$), constructed with an overlapping-sphere model. Both packing
fraction, $\phi$, and coordination number, $z$, decrease monotonically  
with microscopic friction $\mu$, for all shapes. However, this decrease is more
dramatic for larger $s$ due to a reduction in the fraction of face-face
contacts with increasing friction.  For flowing grains, the dynamic friction
$\tilde{\mu}$ - the ratio of shear to normal stresses - depends on shape,
microscopic friction and inertial number $I.$ For all shapes, $\tilde{\mu}$
grows from its quasi-static value $\tilde{\mu}_0$ as
$(\tilde{\mu}-\tilde{\mu}_0) = dI^\alpha,$ with different universal behavior
for frictional and frictionless shapes. For frictionless shapes the exponent
$\alpha \approx 0.5$ and prefactor $d \approx 5\tilde{\mu}_0$ while for
frictional shapes $\alpha \approx 1$ and $d$ varies only slightly.  The results
highlight that the flow exponents are universal and are consistent for all the 
shapes simulated here.

\end{abstract} 

\maketitle

Granular materials are ubiquitous in engineering, industrial, and natural
settings. Understanding packing, mechanics, and flow of granular materials like
metallic and polymeric powders or rocks and soils is not only of fundamental
physical interest but also of important practical concern. Significant advances
have been made understanding the nature of granular statics and dynamics
through extensive experimental and computational studies of monodisperse
spheres \cite{vanHecke.2010,Franklin2016handbook}.  

It is well-established that the microscopic particle friction, $\mu$,
strongly influences the stability of static packings of spherical particles,
allowing sphere packs to span the density range from random close packing,
nominally identified with the packing fraction, $\phi_{\rm{rcp}} \approx 0.64$
at $\mu=0$, down to random loose packing, $\phi_{\rm{rlp}} \approx 0.55$ for
$\mu \gtrsim 0.5$ \cite{Torquato.2000, Zhang.2001, Silbert.2002, Jerkins.2008,
  Jin.2010, Farrell.2010, Silbert.2010}.  Correspondingly, the coordination
number $z$, the average number of contacting particles, also exhibits a
continuous decrease from $z_{\rm{rcp}} = 6$ to $z_{\rm{rlp}} \approx 4$
\cite{Zhang.2001, Silbert.2002, Zhang.2005, Silbert.2010}. From a dynamic
view, granular materials similarly express a rich rheology, particularly flows
of dense, cohesionless grains \cite{Pouliquen.2008, Fall.2015,
  Borzsonyi.2017}.  Computer simulations continue to prove useful by providing
further insight into the rheology of granular materials, especially the role
of particle friction \cite{Otsuki.2011, Hurley.2015, Degiuli.2016}, and by
offering ways to test and validate efforts to develop constitutive models
\cite{Kamrin.2014}.

The behavior of aspherical particles is less studied, although it is known
that particle shape has an important role in modifying packing \cite{Zou.1996,
Man.2005, Franklin.2006, Blouwolff.2006, Jaoshvili.2010, Athanassiadis.2014,
Ishan.2014, Roth.2016, Meng.2016, Malmir.2017, Zhao.2017}, flow
\cite{Borzsonyi.2007, Borzsonyi.2013, Wegner.2014, Cwalina.2016}, and quasi-static mechanical properties \cite{Radjai.2012, Sornay.2013, Saussine.2007, Lizcano.2013, Saussine.2009}. At a
practical level, most real particulates are frictional and far from spherical,
from grains of sand and stones to corn kernels and coffee beans. While shape
tends to cause the packing density of frictionless packings to increase with
increasing asphericity, at least until the particle aspect ratio exceeds some
threshold \cite{Donev.2007, Schreck.2010}, there are some hints that despite
differences in shape, frictional nonspherical particles share similarities with
spheres \cite{Kudrolli.2010}.

\begin{figure}[h!]
\vspace{-0.25cm}
\includegraphics[clip,angle=270,width=3.35in]{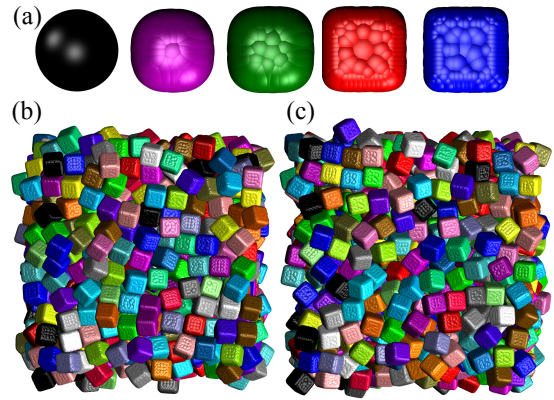}
\vspace{-0.25cm}
\caption{ (a) Superballs created using the overlapping sphere
  algorithm. From left to right: $s=2.0$, $2.5$, $3.0$, $4.5$, and
  $6.0$. (bottom) Static packings for $s=6.0$ superballs with microscopic
  friction (b) $\mu=0.0$ and (c) $\mu = 1.0$. }
\label{f:superballs} 
\end{figure}

To address the role that particle shape plays in influencing
the packing and flow properties of aspherical, frictional materials, we choose
a series of particle shapes that can be systematically controlled. One
particular class of particle shapes that has received attention are
superquadric particles, or superballs \cite{Cleary.2002, Jiao.2009, Jiao.2010,
  Batten.2010, Royer.2015, Audus.2015}, which are defined by the surface
equation:
\begin{equation}
\left|\frac{x}{a}\right|^s+\left|\frac{y}{b}\right|^t+\left|\frac{y}{c}\right|^v
= 1.
\end{equation} 
Here, we restrict our study to shapes with $a=b=c$ where $a$ is the
characteristic particle length, and to a single shape parameter $s=t=v$. These
shapes lie on the spectrum from a sphere of radius $a$ for $s=2,$ to a cube of
side $2a$ for $s=\infty.$ We study the values $s=2.0$, $2.5$, $3.0$, $4.5$,
and $6.0.$ that represent the transition from spherical to cube-like shapes,
as shown in Fig.~\ref{f:superballs} (a).

\begin{figure}[h!]
\includegraphics[trim={0.2in 0 0 0},clip,width=3.2in]{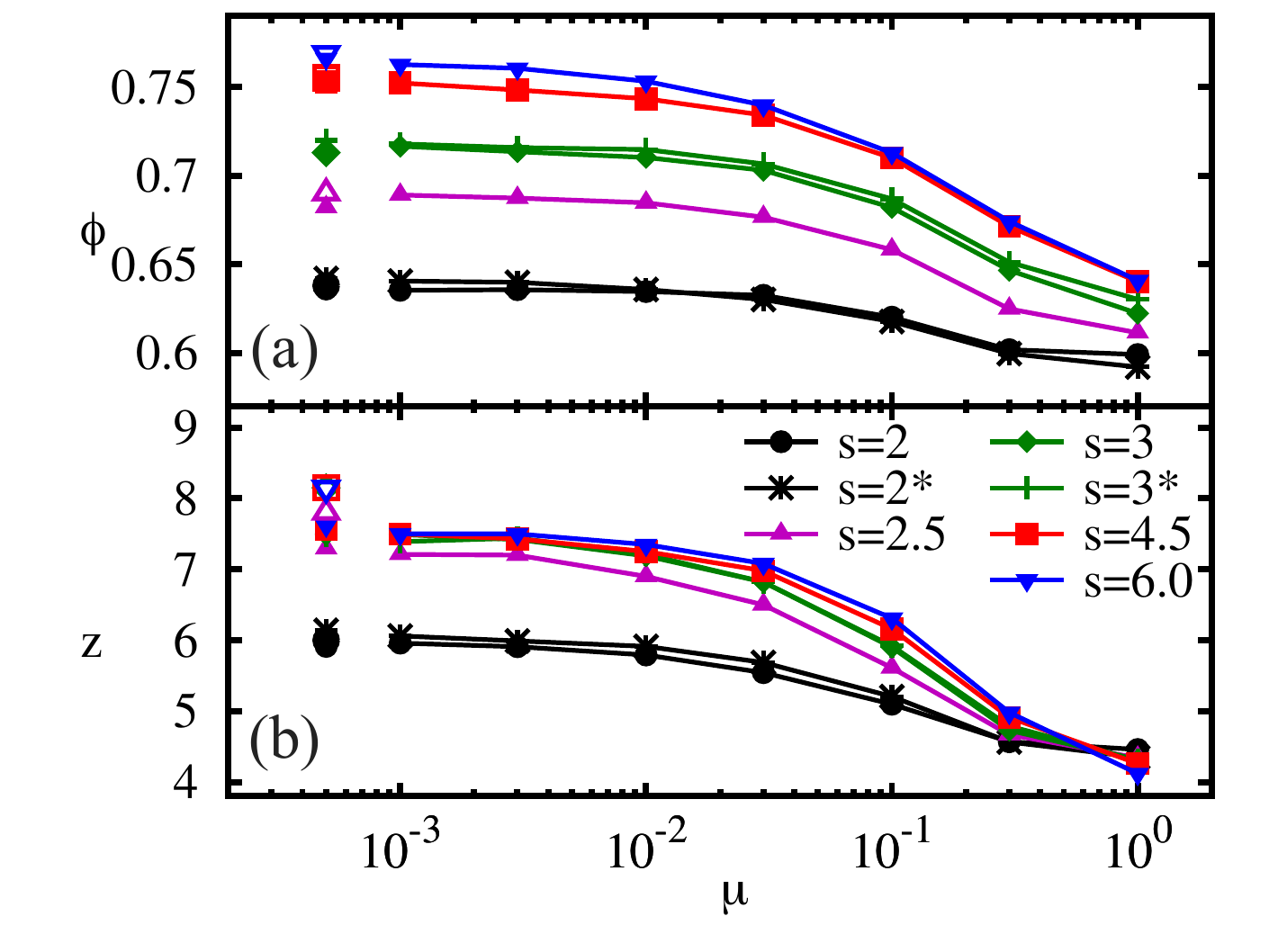}
\caption{(a) Packing fraction, $\phi,$ and (b) coordination number, $z,$
of superball packings over a range in shape parameter, $2 \leq s
\leq 6$, and particle friction coefficient, $0 \leq \mu \leq 1.0$. Symbols
at $5\times10^{-4}$ represent results for $\mu=0$ (solid) and from Jiao et
al.~[\citenum{Jiao.2010}] (open). The * symbols represent bodies of
different fidelity as described in the main text.  }
\label{f:packing-friction} 
\end{figure}

Although granular simulations of frictionless aspherical particles is now a
well-established technique \cite{Nolan.1995, Dziugys.2001, Langston.2004,
  Wouterse.2007, Schreck.2010, Hilton.2010, Abedi.2011, Phillips.2012,
  Azema.2012, Hohner.2012, Guo.2013, Katagiri.2014, Ishan.2014, Dong.2015,
  Zhao.2015}, implementation of the contact mechanics between individual rigid
bodies can be cumbersome, especially when tracking static friction forces for
the duration that two rigid bodies remain in contact. To overcome
contact-detection issues for arbitrarily-shaped, composite rigid bodies, we
implement a clustered-overlapping sphere algorithm \cite{Latham.2009} to
construct superballs comprised of many component spheres of different sizes.
The overlapping-sphere algorithm efficiently packs spheres of variable
diameter to fill an arbitrary three-dimensional shape with an algorithm
similar to other efforts \cite{Phillips.2012}. Although the overlapped-sphere
representation is not perfect, with small gaps between the spheres and surface
corrugation relative to the ideal analytic shape, we maintain a balance
between the number of spheres used in a shape representation and the fidelity
of the representation by using representations that fill at least 95\% of the
ideal shape volume with as few spheres as possible. For $s = 2.0,$ 2.5, 3.0,
4.5, and 6.0, the rigid bodies contain $n= 1,$ 163, 71, 179, and 229 spheres,
which, respectively, fill 1, 0.9866, 0.9760, 0.9674, and 0.9633 of the ideal
superquadric volume. To test the effect of shape fidelity, we also created
spheres using $n= 73$ by applying the overlapping-sphere algorithm after
placing the first sphere off-center, and a superball with $s=3.0$ using $n =
125$ spheres. These systems are denoted by $s=2^{*}$ and $3^{*}$,
respectively, in Fig.~\ref{f:packing-friction}.

The net force and torque between two contacting rigid bodies are computed from
the set of all forces between each pair of contacting spheres that compose the
two bodies. Spheres interact via an established linear spring-dashpot contact
interaction model \cite{Cundall.1979, Silbert.2001}, with normal (n) and
tangential (t) forces parameterized by spring constants $k_{n,t}$ and damping
factors $\gamma_{n,t}$, respectively. In this work, $k_{n} = 2 k_{t} = 200000$
and $\gamma_{n} = 2 \gamma_{t} = 33.5 \tau^{-1}$, throughout, such that the
coefficient of restitution, $e = 0.84$, where time is normalized by the time
unit $\tau = \sqrt{m/k_{n}}$, with $m$ the characteristic mass of a rigid
body. The microscopic, sphere friction coefficient, $0 \leq \mu \leq 1$,
represent realistic friction values. All lengths in the simulation are scaled
by $a$, the characteristic length. Particle motion was integrated via the
velocity-Verlet algorithm while that of the rigid bodies used the method of
quaternions \cite{allen.1987} within the open-source LAMMPS software package
\cite{Plimpton.1995}. The stiffness of the sphere sets the scale for energy
and stress, therefore stress and pressure are scaled by $k_{n}/a$.

Mechanically stable packings were generated, adapting an isotropic compression
protocol with periodic boundary conditions in all three directions
\cite{Silbert.2010}, close to the limit of marginal stability with a packing
pressure $P \approx 10^{-5}$. Figure \ref{f:packing-friction} shows our results
on the packing fraction (a) and coordination number (b) as a function of microscopic
friction $\mu$ over the range, $ 2 < s < 6$. We also use
Fig.~\ref{f:packing-friction} to illustrate the effectiveness of the
overlapping sphere model implemented here by comparing our data to the results
of hard-particle, event-driven dynamics simulations of Jiao et al.~\cite{Jiao.2010}
(open symbols) for $\mu = 0$ only. While the fidelity of the overlapping
sphere method leads to minor deviations from the ``exact'' ($\mu=0$) results,
packing fraction values deviate within just a few percent between the different
overlapping-sphere representations and the hard-superball simulations
\footnote{Due to the approximate superball representation of the
overlapping-sphere model, the packing fraction values given in
Fig.~\ref{f:packing-friction} have been corrected to account for the volume
occupied by 1000 {\it ideal} superballs.}. As Fig.~\ref{f:packing-friction}
displays, superball packings exhibit similar features to spheres: A monotonic
decrease in the packing fraction $\phi$ and coordination number $z$ with
increasing $\mu$, for all $s$. One striking feature is that in the large-$\mu$
limit, packings of different shapes tend to converge to a similar state. In
other words, the reduction in $\phi$ and $z$ with $\mu$ is more dramatic for
larger $s$, causing frictional superballs with different $s$ to all approach
similar values in $\phi$ and $z$. Similar to previous studies of frictionless
shapes, we also observe that for $s \geq 3$ the $\mu = 0$ contact numbers are
approximately constant \cite{Jiao.2010}.  Indeed, $z$ values for the entire
$\mu$ range are very similar for $s \geq 3$.  Our $\mu = 0$ $z$ values lie about
$7-10\%$ below those of Jiao et al, while our values for $\phi$ are in 
significantly better agreement. \cite{Jiao.2010}.  These differences arise
from several factors including the computational methodology,the shape 
interactions (hard vs soft-sphere), and especially the overlapping sphere 
representation, which leads to overly "rounded" shapes and can lead to multiple
contacts between pairs of shapes.  In particular, the overlapping-sphere representation
works to reduce the contact number for a given $s$ value in the jammed state.  In contrast, our contact number results for $s=2$ and $s=2^*$ agree
within 1-2\% with previous results, suggesting that the representation 
and packing protocol is more important for shapes than for spheres.  The
change in $\phi$ with $\mu$ and $s$ shown in Fig.~\ref{f:packing-friction}
arises from the difference in stability of various contact topologies, such as
face-face, edge-face and corner-face, at different friction values, as
described below. 

\begin{figure}[h!]
\includegraphics[width=3.4in]{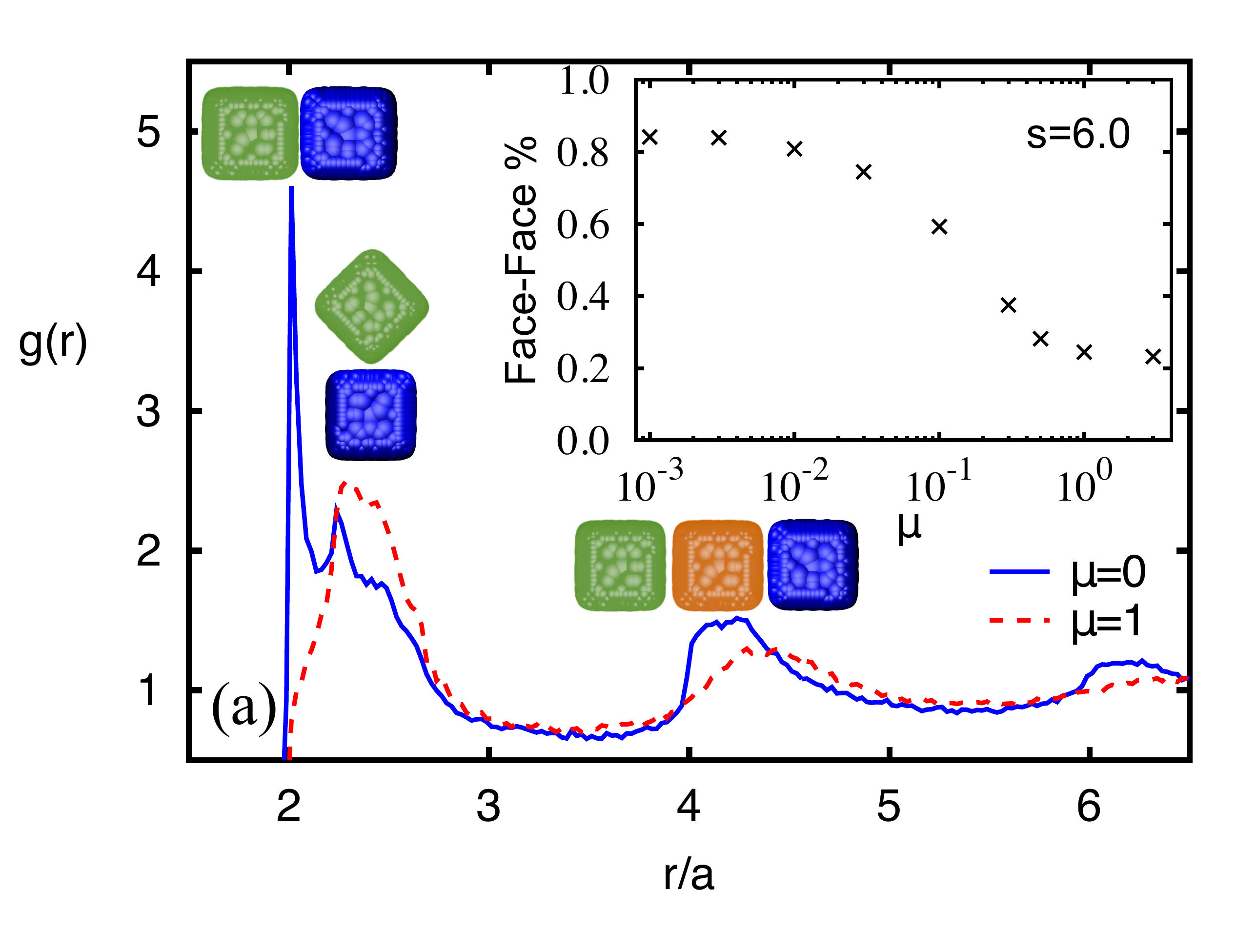}
\includegraphics[clip,width=3.0in]{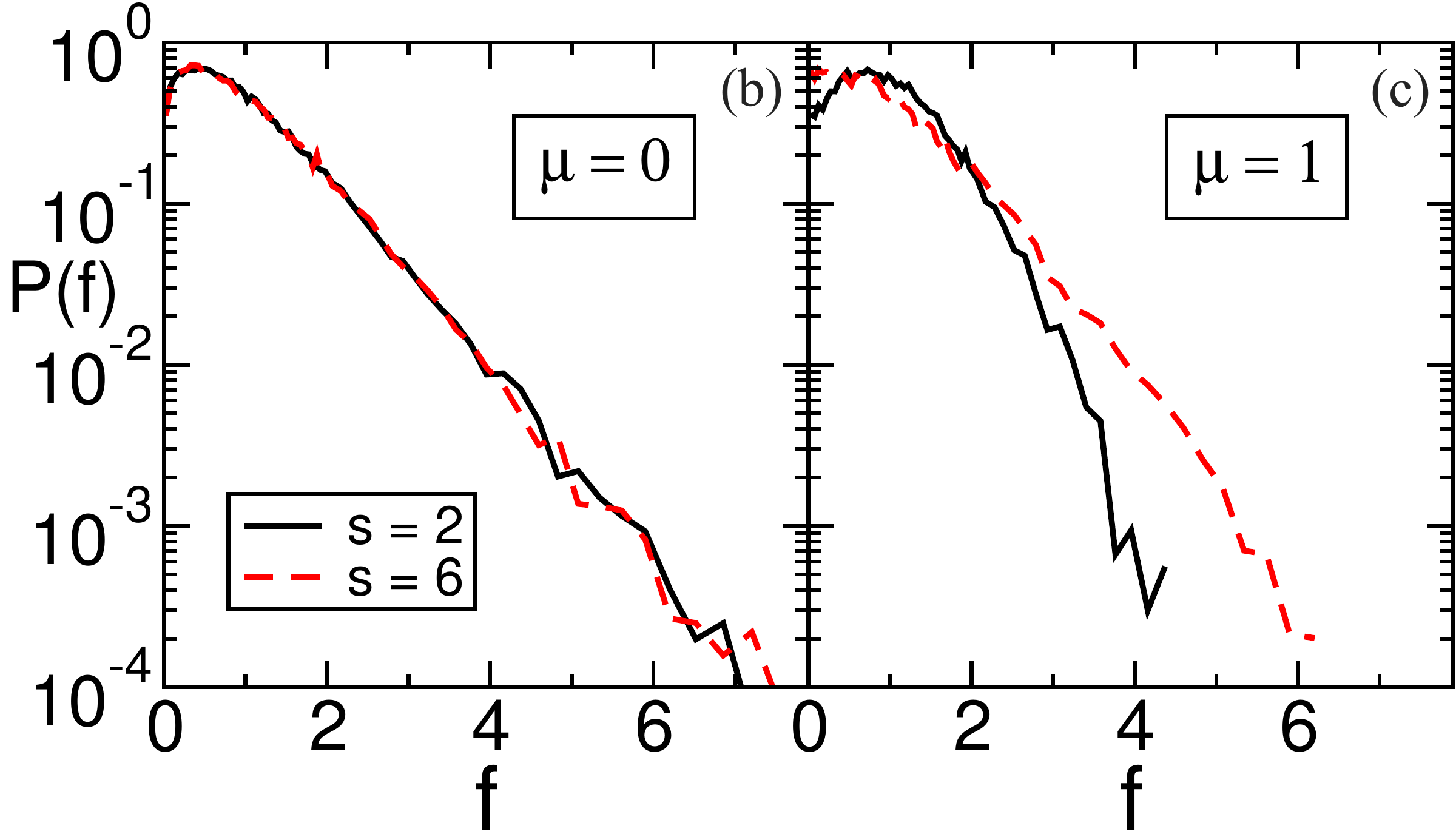}
\caption{ (a) Radial distribution function, $g(r)$, for $s=6.0$ with particle
friction $\mu=0.0$ and $\mu=1.0$. The images reflect likely local structures:
face-face, edge-face, and face-face-face, at the their respective separations,
$r$. (inset) The fraction of particles for $s=6.0$ that have at least one
face-face contact as a function of the surface friction $\mu$.  (bottom) The
distribution $P(f)$, of the normalized, normal contact forces $f$ for $s=2.0$
and $s=6.0$ packings for (b) $\mu=0$ and (c) $\mu=1$.  } \label{f:gofr} \end{figure}

From the radial distribution function $g(r)$, shown in 
Fig.~\ref{f:gofr} (a), there is a distinct shift and broadening of the primary,
nearest-neighbor, first peak for $s=6$ as $\mu$ increases, a feature that is
absent for spheres \cite{Silbert.2002}. For $\mu=0$, the nearest-neighbor peak
represents face-face contacts which leads to efficient packing. The Fig.
\ref{f:gofr} (a) inset shows the dramatic decrease in the fraction of particles
with at least one face-face contact with increasing $\mu$.  While
the presence of face-face contacts has been shown to stabilize packings of
frictionless Platonic solids \cite{Smith2011}, for frictional particles other
contact topologies such as face-edge contacts become more prevalent.  At
$\mu=1.0$, many of these face-face contacts are replaced by local face-edge or
face-corner contacts with increasing $s,$ as surmised from the shift of the
primary peak in $g(r)$.  In addition, the distinctive split second-peak
that is apparent for dense sphere packings is smoothed in the case of
superballs, and broadens with increasing friction.  As a consequence, these
structural dilatational effects cause a decrease in $\phi$ with increasing $s$,
as all shape-packings approach similar values of $\phi$ and $z$ in the large
friction limit. Despite these differences, the distributions of normal contact
forces, $P(f)$, shown in Fig.~\ref{f:gofr} (b) and (c), where we
compare sphere- and cube-packings, suggest that shape has little effect on the
packing mechanical properties for $\mu=0$, indicating similar behavior to two-
dimensional shapes \cite{Radjai.2011}. While at $\mu=1$ subtle differences
such as enhancement of the large-force tail and an increase in the fraction of
smaller forces occur.

\begin{figure}
\includegraphics[trim={0 1in 0 0.7in},clip,width=3.7in]{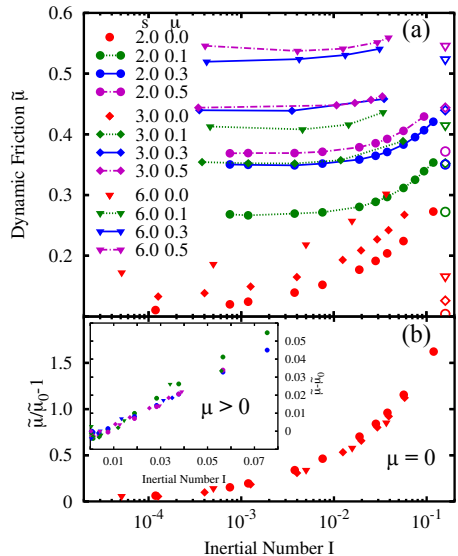}
\caption{Rheology curves for flowing superballs with $s =$ 2.0, 3.0, and
  6.0. (a) Dynamic friction coefficient $\tilde{\mu}$ as a function of the
  inertial number, $I$. At small $I$, all curves approach a constant value.
  Values of $\tilde{\mu}_{0}$ used in (b) are indicated by open symbols at $I
  = 0.16$. (b) The scaled dynamic friction coefficient $\tilde{\mu} /
  \tilde{\mu}_{0} - 1$ for $s=$ 2.0, 3.0, and 6.0 at zero friction.  Inset:
  The shifted dynamic friction coefficient $\tilde{\mu} - \tilde{\mu}_{0}$ 
  for frictional shapes.  }
  \label{f:imu}
\end{figure}

\begin{figure}[!ht]
\includegraphics[clip,width=3.2in]{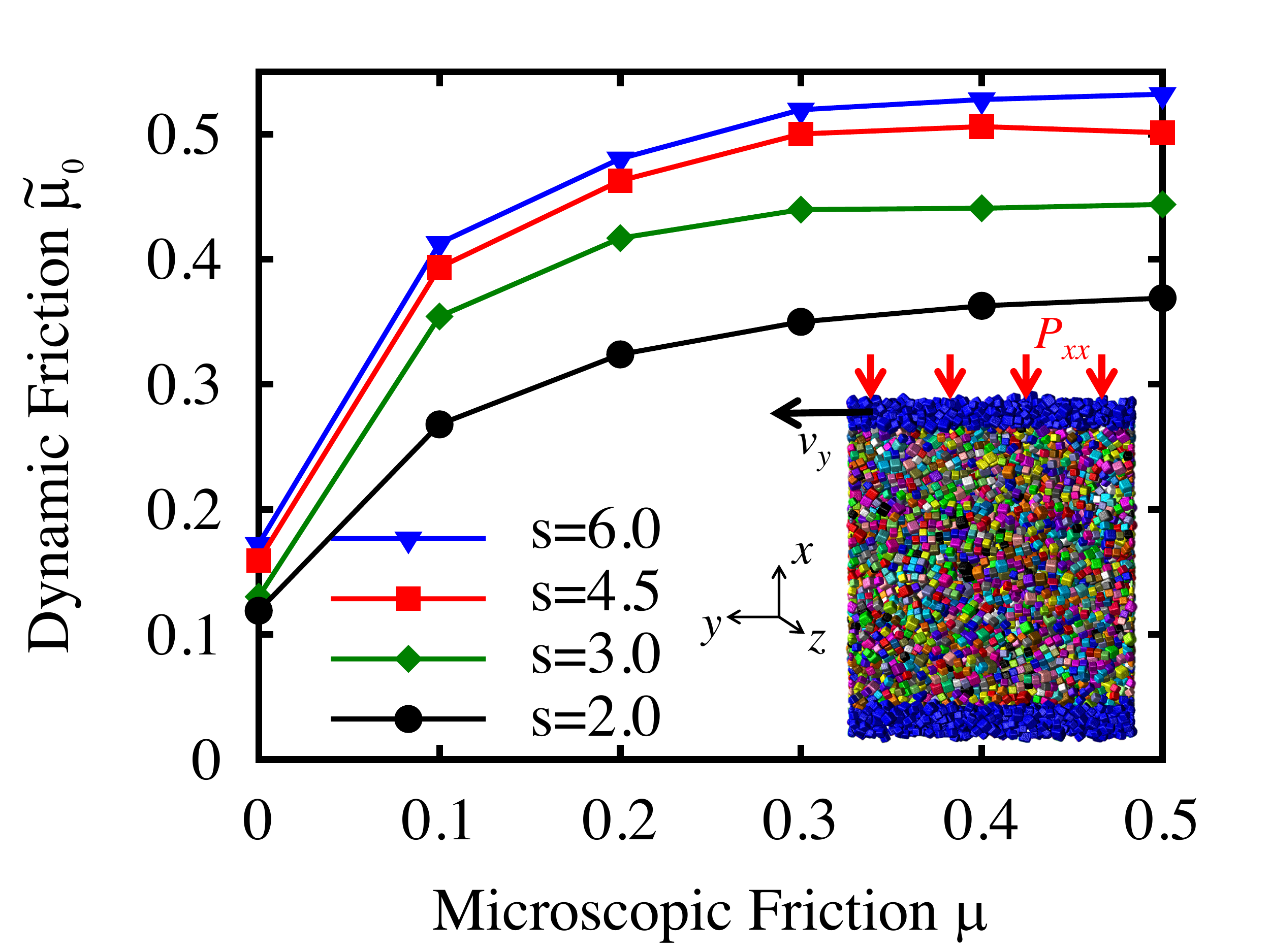} \caption{The dynamic
friction coefficient $\tilde{\mu}_{0}$, in the quasistatic limit ( $I \to 0$),
as a function of the microscopic friction $\mu$, for shape parameter $s =$ 2.0,
3.0, 4.5, and 6.0. Inset: A schematic of the flow sample showing how the wall
velocity and pressure are applied. } \label{f:micro-macro} 
\end{figure}

We now turn our attention to flow. Our flow results are shown in
Figs.~\ref{f:imu} and \ref{f:micro-macro}. Numerous studies of granular flows
\cite{Campbell.2002, daCruz.2005, Otsuki.2011, Kamrin.2014, Hurley.2015,
  Degiuli.2016, Campbell.2016} have highlighted the influence of microscopic
friction on the ratio of the shear stress to the normal stress. It is useful
to think of this ratio as the bulk, macroscopic, \emph{dynamic friction
  coefficient}, or stress anisotropy, $\tilde{\mu}$. This dynamic friction
coefficient scales with inertial number $I$ according to the rheological law,
\begin{equation} \tilde{\mu} = \tilde{\mu}_{0} + d I^{\alpha}, \end{equation}
with $\tilde{\mu}_{0}$ the value in the quasistatic limit $I \to 0$.  The
inertial number $ I = 2\dot{\gamma}a\sqrt{\rho/P_{xx}}$ is a dimensionless
number that depends on the strain rate $\dot{\gamma}$, particle diameter $2a$,
particle density $\rho,$ and confining pressure $P_{xx}$. For spheres, the
power law exponent, $\alpha$, is distinct for frictionless and frictional
particle flows: For $\mu=0$, $\alpha \approx 0.5,$ while for $\mu > 0$, $\alpha
\approx 1.0$ \cite{Bouzid.2015}.

Flowing states contain $N = 6250$ superballs of radius ratio $1:1.4$ and
number ratio $1.4^{3}:1$ to preserve equal volumes of each species and to
avoid ordering during flow.  Initially, dilute samples are compressed
along the $x$-direction by two rigid walls, with the $y$- and $z$-directions
periodic. The sheared system has geometry $L_{x} \approx 90a$, $L_{y} = 70a$,
$L_{z} = 10a$, as shown in the inset to Fig.~\ref{f:micro-macro}. The
$x$-position of the upper wall is pressure-controlled, according to the
equation $\dot{x}_{\mathrm{upp}} = (P(t) - P_{xx})A/\Gamma,$ with damping
parameter $\Gamma = 44 m/\tau,$ $A$ the cross-sectional area, and with a
target normal pressure $P_{xx}=10^{-3}$. Shear flow was imposed by applying a
constant $y$-velocity $v_{\mathrm{upp}}$ to the top wall.  We varied
$v_{\mathrm{upp}}$ to span a range of dimensionless inertial number $I$, from
rapid flow ($I \approx 0.1$), through the inertial regime, down to the
quasistatic limit, $I < 10^{-4}$.

Rheology data over the full range of $I$ shown in Fig.~\ref{f:imu} (a) show
the quantitative dependence of the dynamic friction on shape. The results span
a wide range of inertial number $I$ for several values of surface friction
$\mu$ and shape $s$.  Estimates of the quasistatic friction value
$\tilde{\mu}_0$ for each $s$ and $\mu$ are shown on the right as open symbols
at $I=0.16$ \footnote{Quasistatic dynamic friction values were determined in
two ways: For frictional data, a linear fit was used on the rheology data to
determine the intercept {$\tilde{\mu}_{0}$}, which agreed with the lowest
measured dynamic friction values {$\tilde{\mu} ( I \to 0)$}, to within
{$1.5\%$}. For frictionless particles quasistatic values are chosen to
eliminate curvature at small {$I$} when data are plotted on log-log axes}.  We
discuss the data for frictionless and frictional shapes separately below.

In Fig.~\ref{f:imu} (b) the frictionless data are presented.
For frictionless particles, we measure that all our data follow the same
power law with exponent, $\alpha(\mu=0) \approx 0.5,$ independent of shape.  This
exponent is slightly larger than previous measurements for spheres, circles and pentagons 
\cite{Azema.2018, Roux.2008}.  We note that we do not determine precise exponents within
our data, but rather aim to compare between frictionless and frictional particles.  Further, the
excellent collapse of $\tilde{\mu}(I) / \tilde{\mu}_{0} - 1$ as a function of
$I$ implies that $\tilde{\mu}_0$ and $d$ are proportional, and we estimate that
$d(\mu=0) \approx 5 \tilde{\mu}_{0}$.  These results indicate the universality
of the $\alpha \approx 0.5$ scaling for frictionless particles and that the
dynamic friction of frictionless particles is controlled by the quasi-static
limit.  We note that these results appear indpendent of how the shapes are 
represented.

The inset panel of Fig. \ref{f:imu} (b) contains data for frictional
particles, plotted  with the quasistatic value $\tilde{\mu}_{0}$ subtracted,
$\tilde{\mu} - \tilde{\mu}_{0}$. All data exhibit approximately linear
dependence, indicating $\alpha \approx 1$. Also, all the data have a similar
slope, suggesting that $d$ is approximately independent of particle shape when
microscopic friction is present \footnote{Plotting $\tilde{\mu}(I)/\tilde{\mu}_0
- 1$ for the $\mu > 0$ data gives a poor collapse.}. The dependence of slope on
  microscopic friction $\mu$ cannot be ruled out by our data, which is
consistent with previous results for frictional flow of discs in two dimensions
\cite{Degiuli.2016}.  We speculate that dimensionality and particle shape may
shift the phase diagram proposed previously \cite{Degiuli.2016} while
maintaining the same qualitative features.

Values of the quasistatic limit of the dynamic friction coefficient
$\tilde{\mu}_{0}$ are shown in Fig.~\ref{f:micro-macro} for $s =$ 2.0, 3.0 4.5
and 6.0 over a range of particle microscopic friction $\mu$. The data indicate
that $\tilde{\mu}$ monotonically increases with $s$ and $\mu$, saturating at a
shape-dependent value at large $\mu$.  

In conclusion, we have shown that the static packing fraction and contact
number for superball packings depend on shape parameter $s$ and microscopic
friction $\mu$, yet follow trends similar to that of spheres.  As particles
become more aspherical face-edge and face-corner contacts stabilize at high
friction, replacing the face-face contacts that pack more densely. Results
show that the rheology of aspherical particles shares similarities with
spheres. In particular, the power-law exponent of the $I$-dependence is $\alpha
= 0.5$ for frictionless particles and $\alpha = 1.0$ for frictional particles,
independent of shape.  This result suggest that results previously found in 
two dimensions extend also to three dimensions\cite{Azema.2018}.  Interestingly, 
the distinction between $\mu=0$ and $\mu > 0$ also applies to the quasistatic 
value of the dynamic friction coefficient
$\tilde{\mu}_0$ and the prefactor $d$ with $d\sim \tilde{\mu}_{0}(s)$ for
frictionless particles, and $d$ approximately constant for the frictional
shapes simulated here. These results indicate both common and distinct aspects
of packing and flow between spherical and aspherical particles. On the one
hand, the microscopic properties of grains can be somewhat overlooked when
discussing general qualitative behavior, while on the other, specific bulk
material properties require a more detailed understanding of the constituent
particles.

\section{Acknowledgements}

K.M.S. was supported in part by the National Research Council Associateship
Program at the US Naval Research Laboratory.  This work was performed, in
part, at the Center for Integrated Nanotechnologies, an Office of Science User
Facility operated for the U.S. Department of Energy (DOE) Office of Science.
Sandia National Laboratories is a multi-mission laboratory managed and
operated by National Technology and Engineering Solutions of Sandia, LLC, a
wholly owned subsidiary of Honeywell International, Inc., for the
U.S. Department of Energy's National Nuclear Security Administration under
Contract DE-NA-0003525.
This paper describes objective technical results and analysis. Any subjective views or opinions
that might be expressed in the paper do not necessarily represent the views of the U.S.
Department of Energy or the United States Government.

%

\bibliographystyle{apsrev4-1}

\end{document}